\documentclass[structabstract]{aa}  
\usepackage{natbib}
\usepackage{graphicx}
\usepackage{txfonts}

\usepackage{amssymb}
\usepackage[notref,notcite]{} %showkeys
\usepackage{color}

\def\lesssim{\buildrel < \over {_{\sim}}}
\def\gtrsim{\buildrel > \over {_{\sim}}}

\begin{document}

\title{Broad Balmer line emission and cosmic ray acceleration efficiency in supernova remnant shocks}

\author{G. Morlino\inst{1}\fnmsep\thanks{email: morlino@arcetri.astro.it}, 
            P. Blasi\inst{1,2}, R. Bandiera\inst{1} \and E. Amato\inst{1}
            }
\institute{$^1$INAF/Osservatorio Astrofisico di Arcetri, Largo E. Fermi, 5, 50125, Firenze, Italy \\
	      $^2$INFN/Gran Sasso Science Institute, viale F. Crispi 7, 67100 LÕAquila, Italy
               }

\date{Received 1 June 2013; Accepted 27 July 2013}

% \abstract{}{}{}{}{} 
% 5 {} token are mandatory
 
  \abstract
  % context heading (optional), leave it empty if necessary  
   {Balmer emission may be a powerful diagnostic tool to test the paradigm of cosmic ray (CR) acceleration in young supernova remnant (SNR) shocks. The width of the broad Balmer line is a direct indicator of the downstream plasma temperature. In case of efficient particle acceleration an appreciable fraction of the total kinetic energy of the plasma is channeled into CRs, therefore the downstream temperature decreases and so does the broad Balmer line width. This width also depends on the level of thermal equilibration between ions and neutral hydrogen atoms in the downstream. Since in general in young SNR shocks only a few charge exchange (CE) reactions occur before ionization, equilibration between ions and neutrals is not reached, and a kinetic description of the neutrals is required in order to properly compute Balmer emission.}
  % aims heading (mandatory)
   {We provide a method for the calculation of Balmer emission using a self-consistent description of the shock structure in the presence of neutrals and CRs, also accounting for the non-Maxwellian distribution of neutrals.}
  % methods heading (mandatory)
   {We use a recently developed semi-analytical approach, where neutral particles, ionized plasma, accelerated particles and magnetic fields are all coupled together through the mass, momentum and energy flux conservation equations. The  distribution of neutrals is obtained from the full Boltzmann equation in velocity space, coupled to Maxwellian ions through ionization and CE processes. The computation is also improved with respect to previous work thanks to a better approximation for the atomic interaction rates.}
  % results heading (mandatory)
{We find that for shock speeds $\gtrsim 2500$ km/s the distribution of broad neutrals never approaches a Maxwellian and its moments differ from those of the ionized component. These differences reflect into a smaller FWHM than predicted in previous calculations, where thermalization was assumed.}
  % conclusions heading (optional), leave it empty if necessary 
   {The method presented here provides a realistic estimate of particle acceleration efficiency in Balmer dominated shocks.}

\keywords{acceleration of particles -- shock waves -- ISM: cosmic rays -- ISM: supernova remnants}

\authorrunning{Morlino et al.}
\titlerunning{Broad Balmer lines as an indicator of CR acceleration efficiency}
\maketitle
%________________________________________________________________

\section{Introduction}

Optical spectra of a few young SNR shocks consist essentially of Balmer emission. Observations show that the H$\alpha$ line produced in these so-called {\it Balmer dominated shocks} is made of two components \cite[see][for a review]{Heng10} : a narrow line, with a FWHM of few tens of km/s produced by excitation of cold atoms entering the shock region, and a broad line with a FWHM of the order of the shock speed produced by atoms which undergo a CE process before being excited \citep{Chevalier78,Chevalier80}. Measuring the shape of this complex Balmer line allows one to infer information on the conditions around the shock, which in turn are affected by the presence of accelerated particles.
Balmer emission from young SNR shocks can be used to measure the CR acceleration efficiency at such shocks and even to gather information on a CR induced precursor upstream of the shock as predicted by non-linear theories of diffusive shock acceleration.

Efficient CR acceleration at SNR shocks changes the Balmer line emission in (at least) two ways. For CR modified shocks the temperature of the ions downstream of the shock is expected to be lower than in the absence of particle acceleration. This leads to a narrower width of the broad Balmer line. On the other hand, the pressure carried by accelerated particles also produces a CR-induced precursor in which ions can be heated due to both adiabatic heating (the plasma slows down) and interaction with magnetic turbulence, possibly induced by the same accelerated particles. This second effect leads to a broadening of the narrow component of the Balmer line. 

An anomalous width of the narrow Balmer line was first reported as early as 1994 \cite[]{Smith94,Hester94}: a FWHM  ranging from 30 to 50 km/s was detected in several SNRs, implying a pre-shock temperature around 25,000--50,000 K, clearly in excess of the temperature for which hydrogen would stay neutral, in the presence of collisional ionization. This suggests that the line is broadened in some sort of pre-shock region, of a sufficiently small size so as to make collisional ionization unimportant. This is exactly the effect expected from a CR-induced precursor, as was already speculated by \cite{Smith94}. From the theoretical point of view, the first attempt at calculating the broadening of the narrow Balmer lines due to the presence of a CR-induced precursor was done by \cite{Wagner09}, who used a two-fluid model to treat ions and CRs but neglected the dynamical role of neutrals. A different model was proposed by \cite{Raymond11} where momentum and energy transfer between ions and neutrals is included, but both the CR spectrum as well as the profile of the CR-precursor are assumed {\it a priori} rather than calculated. Both works concluded that the observed width of 30-50 km/s can be explained by postulating some level of CR acceleration at the shock. A more detailed calculation, taking into account the mutual interplay between CRs, neutrals, ionized plasma and magnetic turbulence, has been developed by \cite{paperIII}. The conclusion of this latter work is that, as far as the broadening of the narrow line is concerned, the main physical effect to take into account is the damping of magnetic turbulence in the CR precursor, while adiabatic compression alone is ineffective.

A smaller FWHM of the broad Balmer line as the one expected from the presence of CRs \cite[]{vanAdel08}, has been inferred in two SNRs: RCW 86 \cite[]{Helder09}, inside our own Galaxy, and SNR 0509--67.5 \cite[]{Helder10} in the LMC.
In the first case, the authors combined proper motion measurements of non-thermal X-ray filaments observed by {\it Chandra} and optical measurements, by VLT, of the width of the broad H$\alpha$ component from the same region, showing that the temperature behind the shock is lower than expected based on Rankine-Hugoniot relations. The conclusion was that a sizable fraction of the energy ($\geq 50\%$) is being channeled into CRs. In fact this conclusion has been somehow retracted by the same authors in a recent paper \citep{Helder13} where they use new measurements of the proper motion of optical filaments, which imply a lower shock velocity with respect to what was inferred from the X-ray proper motion. Using the new measurements, the CR acceleration efficiency that the authors estimate is lower and depends on the assumed distance of the remnant, being compatible with zero for $d\lesssim 2.5$ kpc.
In the second remnant, SNR 0509--67.5, \cite[]{Helder10} estimate that the CR pressure behind the SW region of the shock is $> 15\%$ of the total post-shock pressure.

In both these papers, the CR acceleration efficiency was estimated through a relation between the FWHM of the broad line and the shock velocity found by \cite{vanAdel08} (hereafter vAHMR) using the formalism developed in \cite{Heng07}. Such relation was based on the assumption that after three CE reactions the distributions of atoms and ions reach thermal equilibrium (namely same temperature and same bulk velocity). Here we use a kinetic approach to show that this condition is not fulfilled and this fact substantially changes the expected width of the broad Balmer line. 

An accurate calculation of the Balmer emission requires a proper description of the distribution function of the neutral particles in the shock region. Neutrals may suffer a number of CE events before being ionized, but in general they do not thermalize with ions, hence the neutral distribution function can only be determined using the full Boltzmann equation. Moreover, when CRs are also present, the shock structure is determined by the mutual interaction between ions, neutrals, CRs and magnetic field, which are all coupled together through different processes, resulting in a highly non-linear system. In a series of three papers \cite[][hereafter Paper I, II and III]{paperI,paperII,paperIII}, we developed a semi-analytical method for the description of the diffusive shock acceleration process in the presence of neutral particles. Our method also provides estimates of the Balmer emission produced by excited Hydrogen atoms in the shock region. 

In the present paper we use the previously developed formalism to assess the relation between broad Balmer line width, shock speed and CR acceleration efficiency. Compared with our previous work, we improved our calculations by implementing a better approximation of the ion-neutral scattering rate. We show that, for shock speed $\gtrsim 2500$ km/s, the neutrals experience just a few CE events before ionization. As a consequence, the assumption that the neutral distribution function converges to the ion distribution function (quasi-Maxwellian assumption) is a bad approximation which leads to overestimating the width of the broad line by a factor up to $\sim20\%$. When measuring the Balmer line width in real sources, one can then easily obtain unrealistically large values for the CR acceleration efficiency, simply as a result from an erroneous prediction of the line width for given shock velocity. We find that for an acceleration efficiency of $\sim 20\%$, the FWHM of the broad Balmer line only decreases by $10\%$ with respect to the case in which CRs are absent. Being the effect so modest, it is clear that if one erroneously expects, based on the assumption of equilibration between ions and neutrals, a very broad line for a given shock velocity, a very large acceleration efficiency ($>50\%$) could be easily inferred. 

In \S~\ref{sec:beta} we summarize the calculations of the neutral distribution function and discuss the improvements we implemented with respect to our previous work. In \S~\ref{sec:res} we discuss our results concerning the distribution of neutrals and the FWHM of the broad Balmer line, comparing them with those in the literature. We conclude in \S~\ref{sec:conc}.

\section{Improved calculation of atomic rates} \label{sec:beta}

We consider a plane-parallel shock wave propagating in a partially ionized proton-electron plasma with velocity $V_{sh}$ along the $z$ direction. The ionization fraction is fixed at upstream infinity (ISM) where ions and neutrals are assumed to be in thermal equilibrium with each other. The shock structure is determined by the interaction of CRs and neutrals with the background plasma, as discussed in detail in our Paper III. In this section we limit ourselves with summarizing the computational method we use to determine the distribution function of neutral atoms in phase space, and illustrate an improved approximation for the ion-neutral scattering rates.

The basic equations that describe the behaviour of neutral hydrogen have been discussed in our Paper I: neutral hydrogen atoms interact with protons through CE and ionization and with electrons through ionization only. The hydrogen distribution function, $f_{N}(\vec v,z)$ satisfies the stationary Boltzmann equation
\begin{equation} \label{eq:vlasov}
v_z \frac{\partial f_{N}(\vec v, z)}{\partial z} = \beta_{N} f_{i}(\vec v, z)  -
        \left[ \beta_{i} + \beta_e \right] f_{N}(\vec v, z) \,,
\end{equation}
where $z$ is the distance from the shock (which is located at the origin, $z=0$), $v_z$ is the velocity component along the $z$-axis and the electron and proton distribution functions, $f_e(\vec v,z)$ and $f_i(\vec v,z)$, are assumed to be Maxwellian at each position.

The collision terms, $\beta_k f_l$, describe the interaction (due to CE and/or ionization) between the species $k$ and $l$. The interaction rate $\beta_k$ can be formally written as
\begin{equation} \label{eq:beta_k}
\beta_k (\vec v,z) = \int d{\vec w} \, v_{rel} \, \sigma(\vec v_{rel})
                  f_{k}(\vec w,z) \,,
\end{equation}
where $v_{rel} = |\vec v- \vec w|$ is the relative velocity and $\sigma$ is the cross section for the relevant interaction process. More precisely, $\beta_N$ is the rate of CE of an ion that becomes a neutral, $\beta_i$ is the rate of CE plus ionization of a neutral due to collisions with protons, while $\beta_e$ is the ionization rate of neutrals due to collisions with electrons. A full description of the cross sections used in the calculations can be found in our Paper~II.

Since the dynamics of electrons is not well known, we parametrize the electron temperature as a given fraction of the proton temperature and we investigate the dependence of the results on such a parameter. In particular we define $\beta_{\rm down}\equiv T_e/T_p$ in the region downstream of the shock and a similar parameter upstream of the shock.

A semi-analytical method for the solution of Eq.~(\ref{eq:vlasov}) has been discussed in Paper I, where we showed how to calculate $f_{N}$ starting from the distribution of protons and electrons. The method is based on decomposing the distribution function $f_{N}$ at each location, upstream and downstream, as the sum of the neutrals that have suffered 0, 1, 2, ..., $k$ CE processes. Each distribution function is named $f_{N}^{(k)}$, and clearly $f_{N}(z,\vec v) = \sum_{k=0}^{\infty} f_{N}^{(k)}(z,\vec v)$. Using this formalism, Eq.~(\ref{eq:vlasov}) can be rewritten as a set of $k+1$ equations, one for each component:
\begin{equation} \label{eq:vlasov_0}
v_z \partial_z f_{N}^{(0)} = -\left[ \beta_{i} + \beta_e \right] f_{N}^{(0)} 
     \quad {\rm for} \; k=0 \,, \quad {\rm and}
\end{equation}
\begin{equation} \label{eq:vlasov_k}
v_z \partial_z f_{N}^{(k)} = \beta_{N}^{(k-1)}
     f_{i} - \left[ \beta_{i} + \beta_e \right] f_{N}^{(k)} 
     \quad {\rm for} \; k>0 \,.
\end{equation}

Each one of these equations is linear, and the solution for $f_{N}^{(k)}$ can be obtained from $f_{N}^{(k-1)}$. The boundary conditions are imposed at upstream infinity where ions and neutrals are assumed to start with the same bulk velocity and temperature, so that CE occurs at equilibrium and the distributions do not change: the resulting boundary conditions are $f_{N}^{(0)}(-\infty,\vec v) = (n_{N,0}/n_{i,0}) f_i(-\infty,\vec v)$ and $f_{N}^{(k>0)}(-\infty,\vec v)= 0$. We can write the solution of these equations in the following implicit form:
\begin{equation}
f_{N}^{(0)} (z,v_{\parallel},v_{\perp}) = f_{N}^{(0)} (-\infty,
v_{\parallel}, v_{\perp}) \, \exp\left[-\int_{-\infty}^{z} \frac{d
z'}{v_{\parallel}} \beta_{i} (z',v_{\parallel},v_{\perp})\right] \,,
\end{equation}
for $k=0$, and 
\begin{equation}
f_N^{(k)}= - \int_{z}^{+\infty} \frac{dz'}{v_{\parallel}} \beta_{N}^{(k-1)}
f_{i} \exp \left[ \int_{z}^{z'} \frac{dz''}{v_{\parallel}} \beta_{i} \right],
~~~v_{\parallel}<0,
\label{eq:fnk<}
\end{equation}
and
\begin{equation}
f_N^{(k)}= \int_{-\infty}^{z} \frac{dz'}{v_{\parallel}} \beta_{N}^{(k-1)}
f_{i} \exp \left[ \int_{z}^{z'} \frac{dz''}{v_{\parallel}} \beta_{i} \right],
~~~v_{\parallel}>0,
\label{eq:fnk>}
\end{equation}
for $k>0$. Notice that $v_{||}=v_z$ is the velocity component along $z$ while $v_{\perp}$ is orthogonal to $z$.

A good approximation to the total distribution function of neutrals is obtained by adding a sufficiently large number of $f_{N}^{(k)}$. How large the number is determined by the physical scale of the problem. A rough estimate can be obtained by comparing the lengthscale for CE with the maximum between the ionization length-scale and the precursor length. In general, for typical values of the ambient parameters, the longest scale, and hence the one relevant for this comparison, is the size of the CR-induced precursor.

From the computational point of view, the most expensive part is the calculation of the interaction rates $\beta_{N}^{(k)}$ from Eq.~(\ref{eq:beta_k}). In order to reduce the computational time, in Paper~I we used an approximated expression, originally proposed by \cite{Pauls95}, which is based on two assumptions: $i)$ the partial distribution functions $f_{N}^{(k)}$ are assumed to be Maxwellian and $ii)$ the cross section is assumed to be constant (or slowly changing as a function of the relative velocity). Under these assumptions the interaction rates can be expressed as follows:
\begin{equation} \label{eq:beta_old}
\beta_{N}^{(k)}(z,v_{||},v_{\perp}) = m_{p} n_{N}^{(k)} (z) \, \sigma (U_{*}) U_{*},
\end{equation}
where $U_{*}=\sqrt{\frac{4}{\pi} {v_{th,N}^{(k)}}^{2} + (v_{bulk,N}^{(k)} - v_{\parallel})^{2} +
v_{\perp}^{2}}$ and $\sigma(U^*)$ is the relevant cross section calculated at $U^*$. Here $n_{N}^{(k)}$ is the numerical density of particles in the distribution function $f_N^{(k)}$, while $v_{bulk,N}^{(k)}$ and $v_{th,N}^{(k)}$ are the moments of the same distribution corresponding to bulk and thermal speed. 

In our calculations the $f_{N}^{(k)}$ are neither assumed to be Maxwellian nor found to be Maxwellian, and most of the difference between the results presented here and previous literature arises exactly from this fact, namely our more realistic treatment of the neutral distribution in phase space. However, we find that the scattering rates can be computed in an accurate way through the following procedure, which allows us to save a large amount of computation time: we compute the moments of the $f_{N}^{(k)}$ and then introduce an equivalent Maxwellian, with the same moments, only to the purpose of calculating the scattering rates. We checked that this procedure does not introduce any critical aspect in the calculations and therefore retain this part of the computation here as well. On the other hand, we here relax the assumption that the cross sections are weak functions of the relative velocity. Following \cite{Chevalier80} the three dimensional integral in Eq.~(\ref{eq:beta_k}) can be reduced to only one dimension with a change of variable. The final result can be written as:
\begin{equation} \label{eq:beta_new}
  \beta_{N}^{(k)}(v,v_{th},z) = n_N^{(k)}(z) \int_0^{\infty} dv_r \frac{\sigma(v_r) v_r^2}{v v_{th} \sqrt{\pi}} 
  \left[ e^{-\left(\frac{v-v_r}{v_{th}} \right)^2} - e^{-\left(\frac{v+v_r}{v_{th}} \right)^2} \right] \,,
\end{equation}
where now $\sigma$ is an arbitrary function of the relative velocity $v_{r}$.
The integration in Eq.~(\ref{eq:beta_new}) is performed numerically only once and for each cross section involved in the calculation, on a two dimensional grid in the variables ($v,v_{th}$). Using Eq.~(\ref{eq:beta_new}) rather than Eq.~(\ref{eq:beta_old}) the value of $\beta_N^{(k)}$ is more accurate by $\sim 10\%$. At the same time, this improved version of the calculation is also found to be computationally faster by roughly 20\%.

\section{Results} \label{sec:res}

The most important result of this work is that for shocks moving with speed larger than $\sim 2500$ km/s, the neutral distribution function does not reach thermal equilibrium with the ions distribution, not even after many CE reactions. As a consequence, we predict widths of the broad Balmer line appreciably smaller than estimated in previous calculations, where the assumption was made that ions and neutrals would thermalize after three CE reactions (vAHMR). The implications for the estimates of the CR acceleration efficiency are rather dramatic and will be discussed below.

As a benchmark case, we consider a shock with speed $V_{sh}= 4000$ km/s, total upstream density $1\, \rm cm^{-3}$ and ionization fraction $x_{\rm  ion}= 50\%$. With these values for the parameters, the typical interaction length for both CE and ionization is $\sim 10^{16}$ cm. For this specific case, we assume, as may be expected for a collisionless shock, that the background electrons and protons are not in thermal equilibrium. We take $\beta_{\rm down}= 0.01$, although this assumption is not crucial for our conclusions. We first consider a case with no particle acceleration. 

\begin{figure}
\begin{center}
{\includegraphics[width=\linewidth]{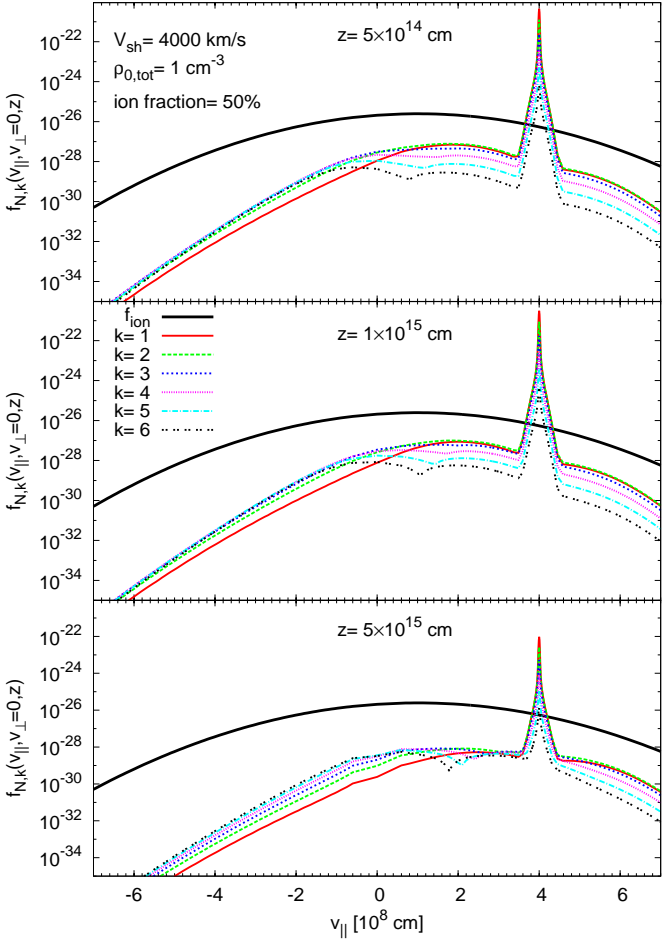}}
\caption{Partial distribution functions of neutrals, $f_{N}^{(k)}$ for $k$ ranging from 1 to 6, as a function of $v_{||}$ and for $v_{\perp}=0$. Different panels show the results for different locations downstream: $z=5 \times 10^{14}$cm (top), $10^{15}$ cm (middle) and $5\times 10^{15}$ cm (bottom). For comparison, the thick solid lines show the downstream distribution of ions.}
\label{fig:fNk}
\end{center}
\end{figure}

In Fig.~\ref{fig:fNk} we plot the partial distribution functions $f_N^{(k)}$ for $k=1-6$ at three different locations downstream, as a function of $v_{||}$ and for $v_\perp=0$. For comparison the solid thick line shows the downstream distribution function of ions. Each $f_N^{(k)}$ includes three different contributions: a broad component, emitted by neutrals that suffered at least one CE downstream, an intermediate component, produced by atoms resulting from CE in the neutral return flux precursor upstream (Paper I), and finally a narrow component, emitted by truly cold ($T\sim 10^{4}$ K) atoms. As far as the atoms responsible for the broad component are concerned, we clearly see that even for $k=6$, namely after six CE events (basically all in the downstream in this case), their distribution does not approach a Maxwellian, and one can easily judge that its peak is still at a different velocity with respect to that of ions. We can make this argument more quantitative by calculating the moments of the distributions.

\begin{figure}
\begin{center}
{\includegraphics[width=\linewidth]{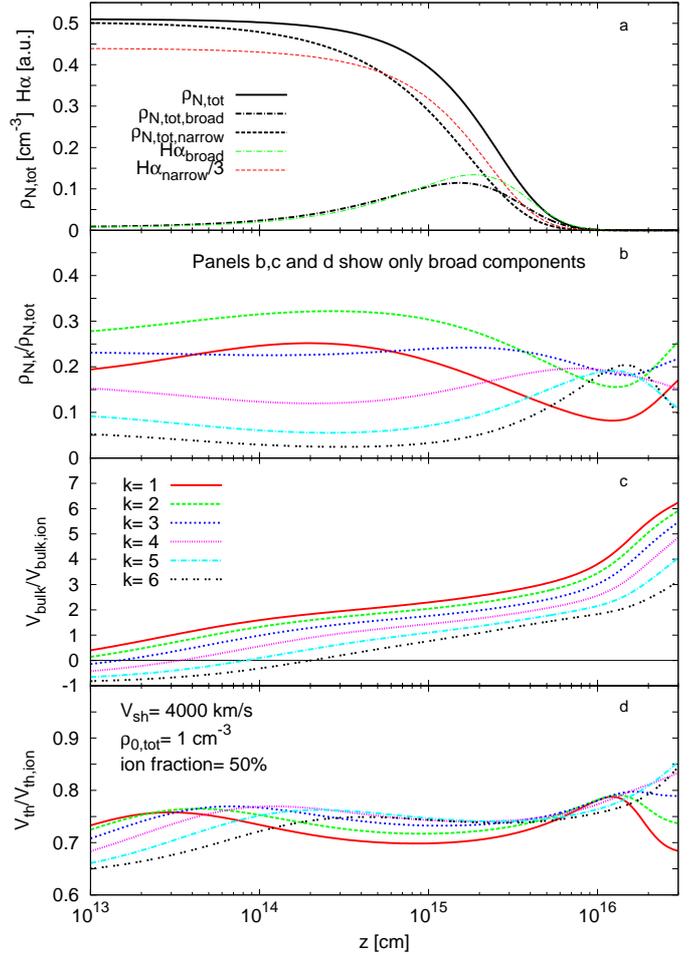}}
\caption{Panel $a$: total density of broad (dashed) and narrow neutrals (dot-dashed) and the sum of the two (solid line). The thin lines show the Balmer emission of narrow (dashed) and broad (dot-dashed) components (the first has been divided by 3). 
Panel $b$: density of the partial component of broad neutrals normalized to the total density of broad neutrals, for k= 1 to 6.
Panel $c$: bulk speed of the partial component of broad neutrals normalized to the bulk speed of ions. Panel $d$: thermal speed of the partial component of broad neutrals normalized to the thermal speed of ions. Notice that the labels in panel $c$ also refer to panels $b$ and $d$. 
}
\label{fig:moments}
\end{center}
\end{figure}

In Fig.~\ref{fig:moments} we plot several quantities related to the neutral distribution as a function of the position downstream, for the same benchmark case. Panels $(b)$, $(c)$ and $(d)$ show, for each $f_N^{(k)}$, the behavior of density, bulk speed and ``equivalent" thermal speed of the broad component only (the bulk and thermal velocities are normalized to the same quantities as computed for the ions' distribution function). The region in which these parameters are most important is where the bulk of Balmer emission is produced. This region is easily identified from panel $(a)$, which shows the numerical density and the H$\alpha$ emission of both the broad and narrow line emitting components: the bulk of the broad H$\alpha$ component is produced in the region $10^{15}<z<10^{16}$ cm.

It is easy to see from panel $(c)$ that in the region between $10^{15}$ and $10^{16}$ cm, the bulk speed of all $f_N^{(k)}$ is larger than the bulk speed of ions, although $v_{bulk}$ decreases for larger values of $k$. We also note that for $z\lesssim 10^{14}$ cm, $v_{bulk}$ becomes negative which is a consequence of the existence of the neutral return flux (see Paper~I). We see in panel $(d)$ that the thermal speed as calculated for any $f_N^{(k)}$ is smaller than the thermal speed of ions by $\sim (25-30)\%$. These two facts prove that the ionization starts dominating over CE before neutrals and ions may reach thermal equilibrium. 

Finally, from panel $(b)$ of Fig.~\ref{fig:moments} we see that for $z<10^{15}$ cm the partial distribution functions with $k>2$ contribute about half of the total density of broad neutrals, while for $z>10^{15}$ cm their contribution becomes larger than 50\%. In other words the bulk of broad Balmer line emission is dominated by neutrals which have suffered more than two CE events rather than the first two components. This fact is especially important in the comparison of our results with those of \cite{vanAdel08}. Similarly to what we do, the latter authors also separate the neutral distribution function into a sum of partial functions each defined by the number of CE events suffered by the particles. Differently from ours, their partial functions are averaged over the downstream region (there is no description of the upstream region in \cite{vanAdel08}). However, what is most important, and makes the biggest difference between our results and those of vAHMR, is their assumption that after two CE events the neutral distribution function acquires the same moments as $f_{\rm ion}$, namely same temperature and same bulk speed.

In the following, we compare our results with those of vAHMR, and show how their assumptions can easily lead to overestimate the CR energy content at the shock. Before proceeding, however, we want to spend a word of caution on the fact that a direct comparison is actually difficult, because of the differences in the two methods: vAHMR only account for the dynamics in the downstream, while we consider also the dynamics of ions and neutrals in the upstream, hence our $f_N^{(k)}$'s also include particles which undergo one or more CE events in the upstream.
 
\begin{figure}
\begin{center}
{\includegraphics[width=\linewidth]{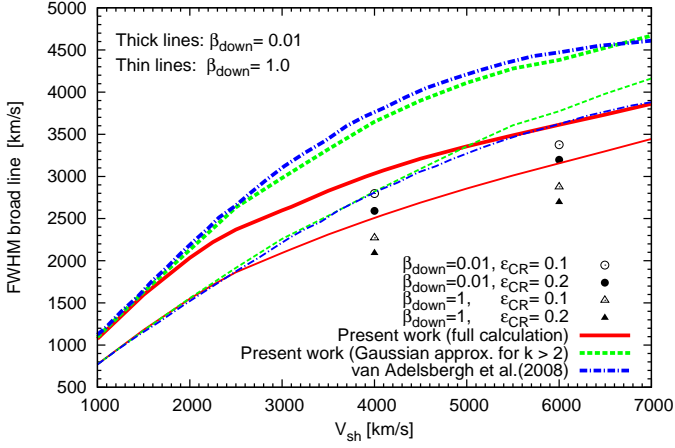}}
\caption{Comparison between the FWHM of the broad Balmer line as calculated in the present work (solid lines) with those calculated by \cite{vanAdel08} (dot-dashed lines). The dashed lines represent the FWHM calculated using our technique, but imposing that the partial distribution functions of broad neutrals are Maxwellian for $k > 2$, with the same temperature and the same bulk velocity as the ions. Dots show the FWHM calculated in the presence of CRs with acceleration efficiency equal to 0.1 and 0.2 and for shock speed of 4000 and 6000 km/s.}
\label{fig:FWHM}
\end{center}
\end{figure}

In Fig.~\ref{fig:FWHM} we plot the FWHM of the broad Balmer line as a function of the shock speed. Three different sets of curves are shown: the solid lines illustrate the results obtained using our full method, while the results of vAHMR are shown as dot-dashed lines. For the sake of comparison of our calculations with those of vAHMR, we also plot the FWHM that would be obtained from our calculations if we made the artificial assumption that all distributions $f_N^{(k)}$ with $k>2$ Maxwellians with the same moments as the ions' distribution (dotted lines). For each set of curves, thick and thin lines represent the non equilibration (NE: $\beta_{\rm down}=0.01$) and full equilibration (FE: $\beta_{\rm down}=1$) cases between protons and electrons, respectively. 

Our results and those of vAHMR coincide for shock speeds $\lesssim 2000$ km s$^{-1}$ for the NE case (or $\lesssim 2500$ km s$^{-1}$ for the FE case), as one may expect based on the fact that for such low velocities ionization becomes important only after a large number of CE pathlengths, thereby making the assumption of thermalization between ions and neutrals a better approximation.

In order to provide an easy way to compare our results with measured values of the FWHM of broad Balmer lines, we looked for a fit to the curves showed in Fig.~\ref{fig:FWHM}. A good fit is provided by the following expression:
\begin{equation} \label{eq:fit}
 W_0(V_{sh},\beta_{\rm down}) = A \sqrt{ \frac{3 \ln 2}{2 (1+\beta_{\rm down})} } \, \frac{V_{sh}}{{\left[ 1 + {(V_{sh}/V_{\rm br})}^3 \right]}^{1/5}} \,,
\end{equation}
which in the limit of low $V_{sh}$ returns the linear dependence of line width on shock speed. The term in the square root gives the slope expected if the neutrals have a Maxwellian distribution with exactly the same temperature as ions \cite[see e.g.][]{Heng10}. The parameter $A$ represents the deviation from this ideal situation and its value changes only slightly between the NE case ($A= 1.126$) and the FE case ($A= 1.096$), while the value of the break velocity $V_{\rm br}$ changes from $V_{\rm br}= 2104$ km s$^{-1}$ (for NE)  to $V_{\rm br}= 3184$ km s$^{-1}$ (for FE).
Eq.~(\ref{eq:fit}) provides a good fit, giving an error less than 2\%, if the shock speed value is in the range between 500 and 7000 km s$^{-1}$.

Aside from the level of equilibration between electrons and protons, the FWHM of the  broad Balmer line can also be affected, at least in principle, by the presence of He. The main effect of He is associated to the fact that fully ionized He by itself can thermalize to a downstream temperature which is $m_{\rm He}/m_{p}$ times the proton temperature. If collisionless processes lead to thermal equilibrium of the two species (Coulomb collisions are too slow to have an effect on the relevant time scales), then the presence of He would reflect in a downstream temperature possibly as high as $\sim 1.3 T_p$. Clearly, in order for this effect to reflect on Balmer emission, thermalization would have to occur on a spatial scale smaller than the ionization length of hydrogen atoms. On the other hand, if  He$^+$ and He$^{++}$ were present, they could play an active role in CE and ionization of H atoms, thereby modifying the distribution function relevant for the Balmer emissivity. 

While in our calculations we neglect the contribution of He, vAHMR take it into account, under the assumption that it is completely neutral upstream. With this assumption, the effects of He can only appear after it is ionized downstream at least once. This fact immediately suggests that the differences shown in Fig.~\ref{fig:FWHM} between the two calculations (ours and vAHMR's) cannot be accounted for by the different treatment of He. The results of the two calculations are in fact coincident for slow shocks. But this is exactly the regime for which the effects of He should be most important, because the ratio between its ionization length and that of hydrogen atoms is smaller for low shock speeds.

The conclusion we draw is that neglecting the presence of He in our calculations cannot be the main source of difference between our results and those of vAHMR. At the same time we can conclude, based on vAHMR's calculations, that He has negligible effects on the system. In fact, their assumption (namely of negligible He ionization upstream) should apply to most (if not all) of the sources of interest, since from Saha's formula one easily sees that having an appreciable fraction of ${\rm He}^+/{\rm He}$ implies a totally negligible fraction of neutral H, if the overall plasma is in thermal equilibrium. A possible exception to this scenario could be produced by the presence of ionizing radiation which preferentially ionizes He.

Let us now turn our attention to the main point we want to address in this paper, namely the estimate of the CR acceleration efficiency from the measurement of the FWHM of the broad Balmer line. It is apparent that our refined treatment of the neutral distribution downstream leads us to expect a FWHM of the broad Balmer line that is smaller than in previous calculations. As a consequence, for a given measured value of the FWHM, we infer a lower CR acceleration efficiency. In fact, in some SNRs, it may turn out to be difficult or even impossible to reach a definite conclusion concerning the presence of CR acceleration.

The theory of Balmer dominated CR modified shocks has been outlined in our Paper III, where we showed in a quantitative way how the FWHM of the broad Balmer line decreases with increasing acceleration efficiency, as expected on a qualitative ground. This approach has been improved further here with a more accurate calculation of the CE and ionization scattering rates, as discussed in \S~\ref{sec:beta}.

The circles and triangles in Fig.~\ref{fig:FWHM} represent the FWHM of the broad Balmer line for the cases of NE ($\beta_{\rm down} = 0.01$) and full equilibration ($\beta_{\rm down} = 1$) respectively, and for acceleration efficiency of 10\% and 20\%. Two values of the shock speed have been considered: 4000 and 6000 km/s. Here we define the acceleration efficiency as $\epsilon_{CR}= P_{CR}/(\rho_{0,\rm tot} V_{sh}^2)$, where $P_{CR}$ is the CR pressure at the shock position and $\rho_{0,\rm tot}$ is the total density at upstream infinity (ions plus neutrals). This definition could be somewhat misleading, in that the shock can accelerate only the ionized component. In this sense, the actual efficiency would be $\epsilon_{CR}^*=P_{CR}/(\rho_{0,\rm ion} V_{sh}^2)= \epsilon_{CR}/x_{\rm ion}$. Since in Fig.~\ref{fig:FWHM} we assumed an ionization fraction of 50\%, we have $\epsilon_{CR}^*=2 \epsilon_{CR}$. 

From Fig.~\ref{fig:FWHM}, it is clear that, given the strong dependence of the implied efficiency on the assumed level of electron-proton thermalization (parametrized by the value of $\beta_{\rm down}$), it is generally difficult to draw firm conclusions about the CR acceleration efficiency unless additional information on $\beta_{\rm down}$ is available. However, if the measured value of the FWHM of the broad line falls below the thin solid line, then one can derive a robust lower limit on the CR acceleration efficiency, provided a reliable measurement of the shock speed is available. 

To first approximation, the FWHM of the broad line decreases linearly with $\epsilon_{CR}$, for $\epsilon_{CR}\lesssim 20\%$. A simple linear fit to the FWHM in the presence of CRs reads:
\begin{equation}
W(V_{sh}, \epsilon_{CR}) \approx W_0(V_{sh}) \left[1-B/W_0(V_{sh}) \, \epsilon_{CR} \right],
\end{equation}
where $W_0(V_{sh})$ is the FWHM in the absence of CR acceleration, and $B= 2000$ km/s. This implies that, in order to obtain a lower limit on the acceleration efficiency, equal, say, to a given value $\bar \epsilon_{CR}$, the accuracy needed in the measurement of $W$ should satisfy the condition $\Delta W/W \lesssim \bar \epsilon_{CR} \, B/W$. Needless to say that the shock velocity as well is needed with decent accuracy: the lower limit should satisfy $\Delta V_{sh}/V_{sh}\lesssim \bar \epsilon_{CR}$. Such constraints can provide hints on the observational requirements needed to determine the presence of CRs in Balmer dominated shocks.

Finally we note that in the absence of accelerated particles, the FWHM does not depend upon the choice of the density and ionization fraction. What does depend on these two parameters is the absolute value of the length scale where the processes of charge exchange and ionization take place. In the presence of accelerated particles, the results depend upon the gas density very weakly, while the dependence of the inferred acceleration efficiencies on the ionization fraction is more substantial but it needs to be assesses on a case-by-case basis. This is the reason why we decided to carry out the calculations of the FWHM for nominal values of the gas density and ionization fraction, since the arguments presented in the paper remain valid in a qualitative way, irrespective of the exact values of parameters.

\section{Conclusions}\label{sec:conc}

The possibility of using optical telescopes to measure the width of the Balmer lines in SNRs and inferring from there the CR acceleration efficiency is one of the big hopes for the future of investigations on the CR origin. The technique is, in a way, calorimetric in that the broad Balmer line contains information about the temperature of the hot proton plasma behind the SNR shock and this temperature is lower when the shock accelerates CRs efficiently. Although this is a simple consequence of energy conservation arguments, a detailed knowledge of the physical phenomena that couple ions and H atoms is needed in order to achieve a quantitative estimate of the CR acceleration efficiency. A complete theory of the dynamics of a Balmer dominated, CR modified shock was recently presented in Paper III, where we used the mathematical formalism introduced in our Paper I. 

In the present paper we have improved on one aspect of that work, by implementing a better calculation of the scattering rates of CE and ionization, the two physical processes responsible for the coupling between ionized and neutral material at the shock. However, the main point we want to stress in this work is that in order to obtain an accurate estimate of the CR acceleration efficiency one has to have a space dependent, kinetic approach to the neutrals, such as the one we put forward in Paper III. To highlight this fact, we have compared our predicted FWHM of the broad Balmer line with the results of vAHMR, widely used in the literature to infer information on CR acceleration from optical measurements. In the work by vAHMR, the simplifying assumption was made that neutrals that suffered more than two CE reactions would reach thermal equilibrium with ions (moreover their calculation does not provide a space dependence of the distribution functions). In the absence of CRs, at 4000 km/s the difference in the FWHM between the two calculations is $\sim 1000$ km/s in the NE case and a few hundred km/s for the case of full equilibration between electrons and protons. This is also the order of magnitude of the reduction in the FWHM expected in the case of CR acceleration, which implies that a proper treatment of the neutral distribution function is absolutely crucial for a reliable quantitative assessment of the CR energy content of SNRs. The assumption of ion-neutral equilibration has been shown to be incorrect for shock speeds above $\sim$2500 km/s. The physical reason for this behavior is in the relative importance between CE and ionization: ions and neutrals tend to reach thermal equilibrium through CE, but unless this happens within a length equal to few times the ionization path length at most, Maxwellian neutrals (with the same temperature as ions) are not what the Balmer line will reflect. For low shock speeds, ionization is inefficient and thermal equilibrium between ions and neutrals is approximately reached, but for high speed shocks ionization dominates upon CE and the neutrals do not thermalize. 

The application of these calculations has far reaching implications: the modeling described in Paper III, supplied with the improvements in terms of accuracy and computation time, provided in the present paper, provides all the necessary tools to determine the dynamical state of a Balmer dominated CR modified shocks and to calculate the shape of the Balmer line (narrow, intermediate and broad) for any CR acceleration efficiency. The whole computation is fast enough to allow for extensive applications to several SNRs of interest. Future combined observations of the overall shape of the Balmer line, as well as of shock velocity and multifrequency spectrum from individual remnants may open the door to a systematic estimate of the CR acceleration efficiency in SNRs. 
In two forthcoming papers we will illustrate the results of our investigation of the Balmer emission from two remnants discussed in the introduction, RCW 86 and SNR 0509--67.5.

\begin{acknowledgements}
We are grateful to D. Caprioli and K. Heng for discussions on the topic. 
This work was partially funded through grant PRIN INAF 2010 and ASTRI. 
\end{acknowledgements}

\end{document}